\documentclass[%
 reprint, 
 amsmath,amssymb,
 aps, prl,
]{revtex4-2}
\usepackage[colorlinks=true, linkcolor=blue, citecolor=blue, urlcolor=blue]{hyperref}
\usepackage{graphicx}
\usepackage{dcolumn}
\usepackage{bm}
\usepackage{subcaption}
\usepackage{ragged2e} 

\begin{document}

\preprint{APS/123-QED}

\title{\textbf{Measuring Electron Energy in Muon-to-Electron Conversion using Holographic Synchrotron Radiation Emission Spectroscopy} 
}%

\author{Nicholas Cutsail}
 \affiliation{Department of Physics, University of California, Berkeley, CA 94720, USA}

\author{Johan Vonk}
 \affiliation{Department of Physics, University of California, Berkeley, CA 94720, USA}
 
\author{Vivek Singh}
\affiliation{Department of Physics, University of California, Berkeley, CA 94720, USA}

\author{Yury G. Kolomensky}\email{yury@physics.berkeley.edu}
 \affiliation{Department of Physics, University of California, Berkeley, CA 94720, USA}
 \affiliation{
 Nuclear Science Division, Lawrence Berkeley National Laboratory, Berkeley, CA 94720, USA}

\date{\today}

\begin{abstract}
The coherent conversion of a muon to an electron in a nuclear field has been one of the most powerful methods for searching for charged lepton flavor violation. Recent advancements have significantly enhanced the sensitivity of $\mu \rightarrow e$ searches, primarily driven by advancements in muon beamline design and low-mass tracking detectors, which afford exceptional momentum resolution. Nevertheless, the performance of these detectors is inherently limited by electron scattering and energy loss within detector materials. We propose a {\em holographic} track reconstruction method that leverages synchrotron radiation emitted by electrons to overcome these inevitable limitations. Similar to cyclotron radiation emission spectroscopy, which has demonstrated outstanding energy resolutions for low-energy electrons, our technique relies on a precision measurement of cyclotron frequency, but in a regime where photons are emitted stochastically and are projected onto a 2-dimensional inner surface of a solenoidal magnet. We outline the concept of such a massless holographic tracker and the feasibility of employing this innovative detection strategy for $\mu \rightarrow e$ conversion. We also address pertinent limitations and challenges inherent to the method.

\end{abstract}

\maketitle


\section{\label{sec:introduction}Introduction}

The Standard Model (SM) of particle physics assumes the fundamental notions of lepton number and flavor conservation~\cite{navas2024review}. Both are accidental symmetries, and a theoretical framework that explains the underlying symmetry leading to these conservation laws still needs to be discovered. The observation of neutrino oscillations, which is possible only if neutrinos have mass, has confirmed lepton flavor violation in the neutral lepton sector and implies that all processes involving lepton flavor violation should manifest at some level in perturbation theory. Therefore, Charged Lepton Flavor Violation (CLFV) remains a subject of intense theoretical and experimental interest that would offer valuable insights into the nature of new physics beyond the SM if observed~\cite{deGouvea:2013zba,Marciano:2008zz,Bernstein:2013hba,Calibbi:2017uvl}. Currently, searches for $\mu^{+} \rightarrow e^{+} \gamma$, $\mu \rightarrow e^{+}e^{-}e^{+}$, and coherent conversion of $\mu^{-} \rightarrow e^{-}$ in the field of a nucleus stand out among all CLFV investigations, offering the most stringent constraints~\cite{MEG:2016leq,SINDRUM:1987nra,SINDRUMII:2006dvw}. These channels have relatively clean final states, consisting only of electrons and photons, and allow an experiment to perform a nearly background-free search using high-intensity muon sources. The essence of our study revolves around the experimental identification of $\mu^{-} \rightarrow e^{-}$ conversion, highlighting its distinctive experimental advantages alongside inherent complexities. It relies on negative muons from a muon beam captured by a target material, forming muonic atoms that cascade down to the ground state. In the SM, muons decay in atomic orbit (DIO) or undergo nuclear muon capture. DIO involves the decay of the bound-state muon to an electron and neutrinos, while in nuclear muon capture the muon combines with a nucleus to produce neutrinos. If $\mu^{-} \rightarrow e^{-}$ conversion occurs, an electron is produced without neutrinos. This electron has a specific energy determined by the muon binding energy ($B_{\mu}(Z)$) and the recoil energy ($R(A)$) of the nucleus:
\begin{equation} 
\label{energy_binding}
E_{\mu e} = m_{\mu}c^2 - B_{\mu}(Z) -R(A)
\end{equation}
where $Z$ and $A$ are the atomic number and mass number of the nuclei, respectively. With only a monoenergetic electron in the final state, $\mu^{-} \rightarrow e^{-}$ conversion is speculated to provide the ultimate sensitivity to the CLFV process in the long term since, unlike the $\mu^{+} \rightarrow e^{+} \gamma$ and $\mu \rightarrow e^{+}e^{-}e^{+}$ processes, it does not suffer from the accidental coincidence background at high muon rates. Additionally, since the muon interacts with quarks in a nucleus, the conversion rate depends on the target nucleus and is model-dependent.

Exceptional experimental progress has been made in the last decade, enabling upcoming experiments such as Mu2e~\cite{Mu2e:2014fns} and COMET~\cite{COMET:2018auw} to improve the sensitivity of the $\mu^{-} \rightarrow e^{-}$ conversion 
by four orders of magnitude, with an additional order of magnitude possible in the next-generation upgrade Mu2e-II~\cite{Mu2e-II:2022blh}. This is enabled by the use of a pulsed beam, a novel muon beamline with graded magnetic field, and state-of-the-art low-mass tracking detectors that let the experiments achieve excellent momentum resolution better than 0.2\%~\cite{Yucel:2022oom,Volkov:2021vmt}. Excellent momentum resolution is critical for higher sensitivity since the DIO electrons constitute an intrinsic background that scales with the muon beam intensity. In the endpoint region, the DIO rate varies
as $\mathrm{(E_{\mu e} - E)^5}$~\cite{Shanker:1981mi,czarnecki2011muon} and can only be suppressed with sufficient momentum resolution for the relativistic electron. 

Current experiments commonly employ low-mass particle tracking detectors within a magnetic field to precisely track the trajectory of the relativistic electron emitted during conversion, facilitating its momentum measurement. However, the momentum resolution of current trackers is inherently limited by fluctuations of the energy loss in the tracking material. Ongoing efforts to further reduce the material budget of these detectors will likely push  current technologies to the limit~\cite{Tsverava:2024ziy,Ambrose2020}. Stochastic energy loss widens the conversion signal, necessitating experiments to integrate over a broader region and resulting in increased DIO background; this re-emphasizes the significance of minimizing energy loss and detector resolution. 

We present the novel idea of using synchrotron radiation (SR) from the emitted electrons for energy reconstruction, eliminating the need for tracking material and minimizing the effect of energy loss on track reconstruction. Our proposed technique is fundamentally based on a non-destructive measurement of the electron's cyclotron frequency by projecting visible SR photons onto a photosensitive detector located on the inner surface of a solenoidal magnet. Precise measurements of times and positions of a set of stochastic photon hits on a two-dimensional cylindrical surface reconstruct the three-dimensional electron trajectories within the solenoidal volume, a technique akin to holography. 
The method of non-destructive radiation spectroscopy, in spirit, is similar to the Project 8 experiment~\cite{Project8:2017nal}, which uses Cyclotron Radiation Emission Spectroscopy (CRES)~\cite{Monreal:2009za} to measure low-energy electrons from $\beta$-decay. However, the implementation of our technique diverges significantly from Project 8, as discussed in the following sections. 

\section{\label{sec:approach} Proposed experimental Approach}

Our proposed technique utilizes the characteristics of SR emitted by the ultrarelativistic electrons ($\mathrm{E_{\mu e} \approx 105}$~MeV, Lorentz factor $\gamma \approx 205$) produced following muon capture in aluminum, the stopping target material used in experiments like Mu2e~\cite{Mu2e:2014fns} and COMET~\cite{COMET:2018auw}. When these high-energy electrons are confined by a uniform axial magnetic field ($\mathbf{B}$), they follow helical trajectories. As charged particles undergo acceleration, they emit SR.

\begin{figure*}[ht!]
    \centering

    \begin{subfigure}[b]{0.48\textwidth}
        \centering
        \includegraphics[width=\textwidth]{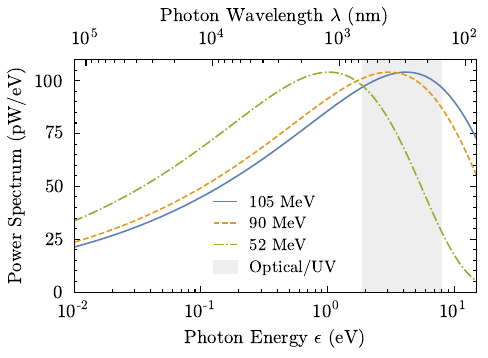}
        \caption{\justifying The 105 MeV case (near conversion endpoint) exhibits peak power in the optical/UV band. Like the 52 MeV Michel decay case, lower energy electrons have lower critical frequencies, contributing less power in this range.}
        \label{fig:powerSpectrum}
    \end{subfigure}
    \hfill
    \begin{subfigure}[b]{0.48\textwidth}
        \centering
        \includegraphics[width=\textwidth]{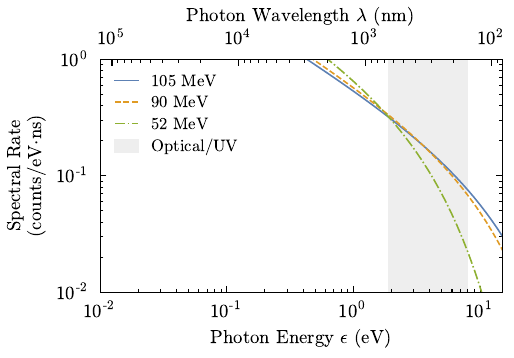}
        \caption{\justifying Expected number of photons from SR. Due to the low number of detectable photons emitted per electron revolution, the quality of the holographic reconstruction is strongly correlated with the total acquired photon statistics.}
        \label{fig:photonNumber}
    \end{subfigure}

    \caption{\justifying Spectral distribution, power and photon rate, for electrons (105, 90, and 52 MeV) in a $B=2\,\mathrm{T}$ field at $\theta=\pi/2$.}
    \label{fig:spectralDistribution}
\end{figure*}

The physics of SR emitted by a single charged particle is well-established and used extensively in scientific research~\cite{Hofmann_2004, Wiedemann2003,SokolovA.A.ArseniiAleksandrovich1986Rfre,Mitsuhashi:2018ree}. Key characteristics relevant to our approach include:
\begin{itemize}
    \item \textbf{High Characteristic Frequency:} The radiation spectrum is effectively continuous, with significant power emitted up to a characteristic critical frequency, $\omega_c$. This frequency is strongly dependent on the electron's energy and the magnetic field strength, scaling proportionally to $\gamma^2 B$~\cite{Hofmann_2004}. Compared to non-relativistic cyclotron radiation, SR frequencies are boosted by relativistic effects (roughly by $\gamma^3$), shifting the emission spectrum significantly. For the $\sim$105~MeV electrons considered and typical magnetic fields (1--3~T), $\omega_c$ falls conveniently within the optical and ultraviolet (UV) range (see Fig.~\ref{fig:spectralDistribution}).
    \item \textbf{High Directionality:} SR is emitted tangentially to the electron's path, highly collimated into a narrow cone (with an opening angle approximately $1/\gamma$) along the instantaneous direction of the electron's motion.
\end{itemize}

These features dictate our experimental strategy. The shift of the dominant radiation frequency into the optical/UV range enables the use of optical photodetectors, significantly altering the detection requirements compared to measuring lower-frequency cyclotron radiation. Furthermore, the highly directional nature of the emitted photons allows us, in principle, to track the electron's helical trajectory within the magnetic field by recording the positions and times of photon hits on the surrounding detector walls (Fig.~\ref{fig:3d-detector}). This concept underpins our Holographic Synchrotron Radiation Emission Spectroscopy (HSRES) approach, which reconstructs the electron’s energy and trajectory by treating the spatial and temporal patterns of synchrotron photons as a holographic imprint of the electron’s motion.

This approach differs significantly from techniques like CRES (e.g., Project 8), which precisely measure a sharp, energy-dependent radio frequency peak from cyclotron motion. In contrast, SR power is spread over a broad range of frequencies (harmonics of the revolution frequency), meaning the energy of individual detected photons does not directly yield the electron's energy. Instead, we propose determining the electron's energy by reconstructing its trajectory from the spatial and temporal patterns of the detected photons. Specifically, we aim to measure the electron's relativistic cyclotron frequency \(\omega=\frac{eB}{m_e\gamma},\) which is inversely proportional to $\gamma$ and thus related to energy, by analyzing the timing and positions of individual photon hits.

Accurate reconstruction of the electron's energy and trajectory relies critically on precise measurement of both the time and position at which each photon is detected. This requirement arises because a single 105~MeV electron emits a limited number of optical photons per revolution (Fig.\ref{fig:photonNumber}), precluding the use of average power measurements for energy determination. Although individual photon emissions are stochastic, the collective distribution of photon hits, resolved in both space and time, encodes the essential information for inferring the electron's path and energy, in a manner analogous to conventional particle tracking systems (Fig.~\ref{fig:projected-track}).

The ultimate energy resolution achievable depends on factors including the electron's observation time within the detector, the total number of detected photons, and the spatial and temporal resolution of the photon sensors. With state-of-the-art technologies such as Large Area Picosecond Photodetectors (LAPPD), capable of timing resolution below 50~ps and spatial resolution on the order of a few mm$^2$~\cite{shin2024advances}, simulations indicate that an intrinsic energy resolution better than ${\Delta E_\text{FWHM}}/{E} < 0.1\%$ is attainable. Such energy resolutions would be unprecedented for conventional tracking detectors or calorimeters.

\subsubsection{Detector Description}

\begin{figure}[ht!]
    \centering
    \includegraphics[width=\columnwidth]{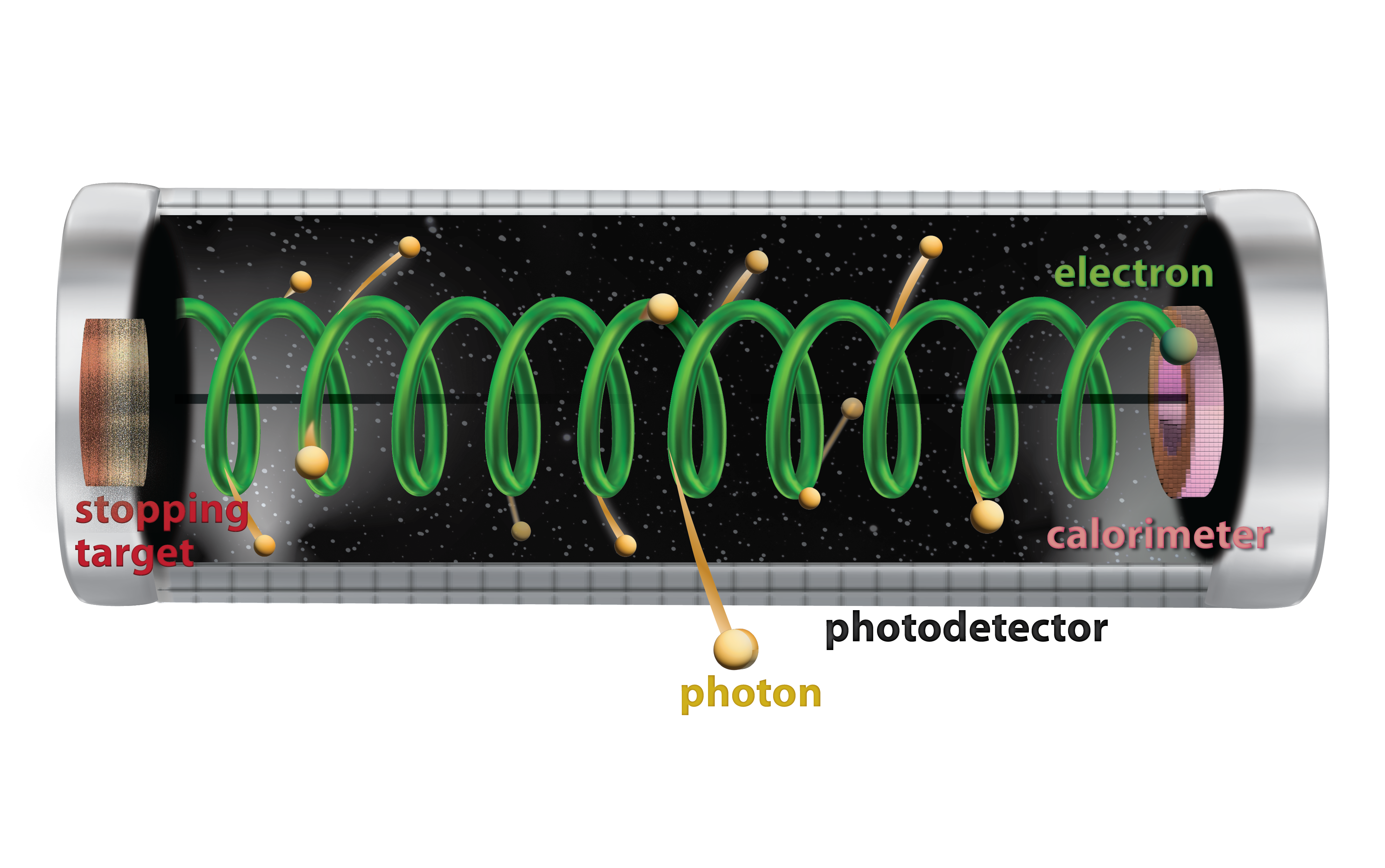}
    \caption{ \justifying An artist's rendition of a conversion electron track inside a hollow photodetector shell. The electron originates at the stopping target and follows a helical path along the magnetic field $\mathbf{B}=B_0 \hat{\mathbf{z}}$, emitting photons tangentially. A calorimeter at the detector end provides approximate timing and position information to seed the trajectory reconstruction. Heavy particles from the accelerator beam or collisions will not emit detectable synchrotron photons due to their low Lorentz factors. Illustration by {\tt SeizetheDesign.com}}
    \label{fig:3d-detector}
\end{figure}

We consider a hypothetical detector setup where a muon beam is stopped and the muon decays in the field of a nucleus, followed by a region of constant magnetic field, where the energy of decay electrons will be measured through their emitted synchrotron radiation (Fig.~\ref{fig:3d-detector}). Inspired by Mu2e~\cite{Mu2e:2022ggl}, we model 34 Aluminum foils (radius 75~mm, spacing 25~mm, thickness 0.1~mm) as a stopping target. 
The stopping target foils are positioned within an axial magnetic field region whose magnitude decreases linearly along the detector axis. This magnetic field gradient acts as a magnetic mirror, significantly improving detection efficiency by redirecting electrons initially emitted backward toward the forward detection region. At higher magnetic fields, the critical pitch angle at which electrons escape upstream increases; however, a larger magnetic gradient ($\Delta B$) simultaneously tends to reduce electron pitch angles by forward-beaming electrons. Because the HSRES method is most sensitive to electrons with larger pitch angles, these two effects compete in determining overall detection efficiency. Through preliminary analysis, we determined that a smooth variation of the magnetic field from $\approx$5~T to 2~T was adequate to balance these competing factors.

The sensitive detection region is a cylindrical volume with a length of 10 meters ($L$) and a radius of 40 centimeters ($R$), immersed in a uniform axial magnetic field. We have chosen a magnetic field strength of $B_0$=2~T; this value balances the increased photon emission rate and critical frequency shift that come with stronger fields, ensuring maximum signal-to-background electron hits for CLFV decay electrons.

The detector's length is a practical compromise, ensuring sufficient photon statistics, as photon hits scale linearly with electron dwell time, while maintaining a manageable size. Its radius is set to fully contain the electron's helical trajectory and minimize the radial dimension. A smaller radius also reduces the spread of photon hits as they reach the cylindrical detector surface.

This volume is surrounded by a cylindrical array of photodetector panels designed to detect photons emitted by electrons traversing the region. The photodetectors are assumed to have state-of-the-art spatial and temporal resolutions of $\sigma_x \approx$~3~mm and $\sigma_t \approx$~50~ps, respectively, and a dark count rate of 1~kHz/cm$^2$, consistent with the current performance of LAPPDs~\cite{shin2024advances}. 
In subsequent analyses, we will systematically vary these resolution parameters around their nominal values to evaluate the sensitivity of our method to detector resolution. The chosen photodetectors have a quantum efficiency profile optimized within the visible wavelength range (approximately 200 to 700~nm).


Finally, at the downstream end of the cylindrical detector, we place a donut-shaped calorimeter to provide initial measurements of electron position, energy, and timing, which serve as seeds for the track reconstruction algorithms. The calorimeter performance is modeled using realistic finite resolutions for energy, position, and timing as reported in Refs.~\cite{atanov2016energy,atanov2015measurement}.

\subsubsection{Simulation}

\begin{figure}[h!]
    \centering
    \includegraphics[width=\columnwidth]{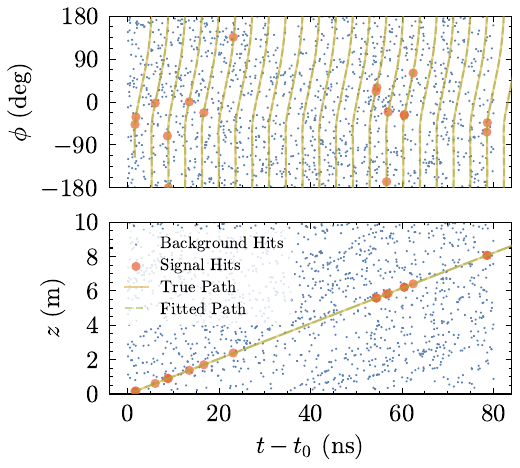}
    \caption{\justifying A simulated conversion electron track is shown. The electron randomly emits photons, which travel in straight paths and hit the detector at an angular coordinate $\phi$, a depth coordinate $z$, and a time $t$. The photon hits are stochastic and sparse in time and are nearly linear in $z$ vs $t$.}
    \label{fig:projected-track}
\end{figure}

We evaluate our method through Monte Carlo (MC) simulations~\cite{Cutsail:2025,Vonk:2025}. Each simulated event window corresponding to a single proton pulse consists of multiple background (DIO) electrons and one potential signal (CLFV) electron. Electrons are independently propagated through the detector, emitting synchrotron radiation (SR) photons, which are then tracked to the photodetector surface. At the detector, the spatial coordinates, $\phi$ (azimuthal angle) and $z$ (longitudinal position), along with the photon detection time
$t$, are recorded. Calorimeter measurements provide information on the electron trajectory at the end of the experimental volume. 

Electron initial conditions, including energy, pitch angle, position, and emission time, are sampled from realistic distributions. Energies follow the spectrum for DIO electrons~\cite{czarnecki2011muon, czarnecki2014michel} or are set at the fixed endpoint for CLFV electrons. Emission times are sampled from the muon lifetime’s exponential decay distribution, truncated to between 700~ns to 1695~ns, matching the Mu2e event window~\cite{Mu2e:2014fns}. Pitch angles account for adiabatic invariance through the magnetic gradient region. Electrons exceeding the critical pitch angle for the magnetic mirror are excluded from further simulation. We note that HSRES is insensitive to heavier particles because the synchrotron photon rate is inversely proportional to mass. As a result, HSRES is less constrained by the beam flash, and the search window may be expanded in time.

Within the sensitive volume, electrons follow helical trajectories at constant energy (neglecting small radiative losses and scattering). Photon emission is modeled as a Poisson process, with photon energies and emission angles sampled from theoretical SR distributions weighted by detector quantum efficiency. Photon hits and calorimeter measurements are smeared according to LAPPD detector resolutions, and detector dark noise is added uniformly.

\subsubsection{Results}

We conducted numerical experiments to investigate how DIO processes affect the energy reconstruction performance of HSRES. Specifically, we studied the effect of varying the average number of DIO electrons per event window, denoted $N_0$. For each fixed value of $N_0$, we generated an ensemble of randomized events and applied pattern recognition and likelihood-based track-fitting procedures~\cite{Vonk:2025,Cutsail:2025} to produce Monte Carlo-estimated reconstructed energy distributions. The chosen range of $N_0$ values encompasses the expected DIO rates from Mu2e Run-I (in both low- and high-intensity modes) and Mu2e-II operational conditions, which we calculate from the stopped-muon and proton intensities in literature~\cite{Mu2e:2022ggl,Mu2e-II:2022blh}, the graded field acceptance efficiency, and the overlap of the 864~ns, exponentially-distributed muon lifetime in Al with the Mu2e search window \cite{Mu2e:2014fns}.

\begin{figure}
\centering
\includegraphics[width=\linewidth]{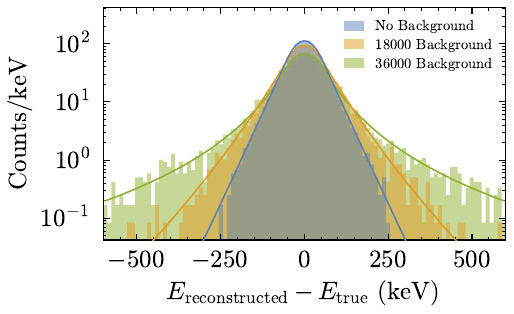}
\caption{\justifying Reconstructed energy distributions for different DIO background levels, fitted with double-sided Crystal Ball functions. Increased backgrounds primarily enlarge distribution tails due to additional random hits passing the $z$-vs-$t$ pattern recognition. The histogram compares the background-free limit to a conservative Mu2e-II intensity scenario ($N_0 = 18,000$) and a higher intensity ($N_0 = 36,000$). Distributions shown correspond to 40\% total efficiency; these tails are significantly reduced at typical operational efficiencies ($\sim15\%$).}
\label{fig:histogram-background}
\end{figure}

\begin{figure}
\centering
\includegraphics[width=\linewidth]{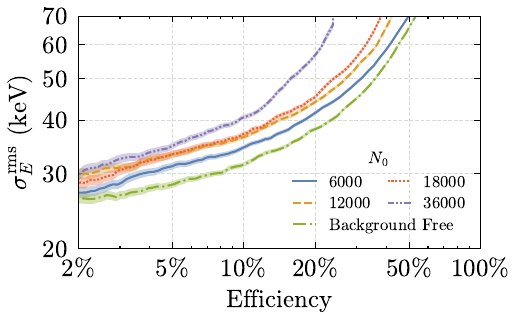}
\caption{\justifying Energy resolution versus total efficiency with 1-$\sigma$ confidence interval for varying DIO backgrounds. Tighter quality cuts improve resolution but reduce efficiency. Higher backgrounds worsen resolution at typical efficiencies. As efficiency approaches zero, resolution asymptotically nears the background-free limit ($\sim$30 keV) due to preferential selection of events with minimally contaminated signal tracks.}
\label{fig:numerical-exp-std-background}
\end{figure}

\begin{figure}
\centering
\includegraphics[width=\linewidth]{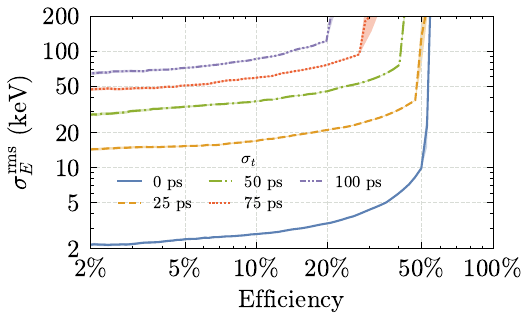}
\caption{\justifying Energy resolution vs. total efficiency with 1-$\sigma$ confidence interval for different photodetector timing resolutions ($\sigma_t$) at fixed DIO background ($N_0=18,000$). Improved timing resolution significantly enhances energy reconstruction, indicating HSRES performance is limited by timing precision rather than synchrotron emission physics.}
\label{fig:numerical-exp-std-time}
\end{figure}

The reconstructed energy distributions follow a double-sided Crystal Ball shape~\cite{Oreglia:1980cs,Gaiser:1982yw} (Fig.~\ref{fig:histogram-background}). With increasing background level, the standard deviation of the reconstructed energy distribution grows, predominantly due to enhanced tails rather than a significant broadening of the Gaussian core. Physically, this occurs because additional DIO electron tracks produce photon hits that overlap with the signal track in the $z$-vs-$t$ plane (Fig.~\ref{fig:projected-track}), causing background hits to pass the pattern recognition stage occasionally. These background hits typically have inconsistent $\phi$ coordinates compared to the signal hits, thus skewing the likelihood minima away from the true parameters and broadening the reconstructed energy distribution.

To comprehensively characterize HSRES performance under varying background conditions, we systematically scanned the space of quality-cut thresholds to quantify the trade-off between reconstruction efficiency and energy resolution. As expected, we observed that tightening quality cuts improves energy resolution at the cost of reduced reconstruction efficiency (Fig.~\ref{fig:numerical-exp-std-background}). Higher background levels yield broader energy resolutions at typical operational efficiencies ($\approx$15\%). However, all energy resolutions asymptotically approach the no-background limit (approximately 30~keV) as efficiency approaches zero. This is because increasingly stringent quality cuts preferentially select events with fewer background hits, irrespective of the underlying DIO rate, driving the resolution toward the intrinsic detector limit. Equivalently, stringent quality cuts remove the tails of the double-sided Crystal Ball distribution, leaving primarily the Gaussian core intact, indicating that the core resolution remains unaffected by moderate background variations.

Having demonstrated the scaling of energy resolution with DIO background levels, we further investigated the intrinsic HSRES resolution by examining how it depends on detector timing and position resolutions relative to the inherent stochastic noise from synchrotron emission. In a second numerical experiment, we varied the photodetector timing resolution ($\sigma_t$) around current state-of-the-art LAPPD values, while keeping other detector parameters fixed and maintaining a representative DIO background for Mu2e-II of $N_0=18,000$.

We found that improved photodetector timing resolution significantly enhances the energy reconstruction performance (Fig.~\ref{fig:numerical-exp-std-time}). This result indicates that current HSRES performance is not fundamentally limited by the stochastic nature of synchrotron emission but rather by the achievable photodetector timing precision. Conversely, varying the photodetector position resolution did not substantially affect the reconstructed energy resolution. Thus, timing resolution emerges as the critical limiting factor, underscoring the necessity of fast photodetectors such as LAPPDs.

\subsection{Summary and Conclusions}
This study presents a novel approach to electron energy reconstruction for the coherent muon-to-electron conversion process, utilizing Holographic Synchrotron Radiation Emission Spectroscopy (HSRES). At the Mu2e-II background scenario ($N_0=18,000$), our simulations indicate that HSRES can achieve an energy resolution of $\sigma_E\text{(core)}\approx50$~keV at a reconstruction efficiency of $25\%$, comparable to the Mu2e-II  efficiency~\cite{Mu2e-II:2022blh}. This projected performance surpasses the current Mu2e or projected Mu2e-II detector capabilities and positions HSRES as a promising candidate for next-generation searches. 

Our analysis demonstrates that the primary limitation to HSRES sensitivity arises from the photodetector timing resolution. Consequently, future improvements in detector timing technologies, such as advancements in LAPPDs, are expected to yield substantial enhancements in achievable energy resolution.

To further refine the accuracy of these projections, several additional effects must be incorporated into future more detailed simulations. For instance, our current analysis neglects small electron energy losses, which could slightly alter the trajectory radii and synchrotron emission frequencies. Given the small fractional magnitude of these losses ($\leq 10^{-6}$), we anticipate only minor adjustments to the fitting templates. Similarly, our assumption of a perfectly uniform magnetic field must be replaced with realistic field maps, necessitating re-parametrization of the fitting templates. However, these refinements are not expected to alter our conclusions significantly. More substantial effects on the energy resolution may arise from non-uniform muon stopping distributions, electron energy straggling due to scattering in the stopping foils, and position correlations at the detector entrance, which are neglected by the adiabatic approximation. 

While our study focused on HSRES energy resolution as applicable for the muon-to-electron conversion, we also note a broadband nature of HSRES. As the spectrum of emitted photons shifts with decreasing electron energy (Fig.~\ref{fig:photonNumber}), reconstruction efficiency decreases but is nonzero down to tens of MeV. Thus, the same detector could be used to search for other CLFV signatures such as $\mu\to eee$, lepton-number violating muon-to-positron conversion, or other exotic new physics signatures~\cite{Hill:2023dym,Fox:2024kda,Bigaran:2025uzn,Hostert:2023gpk}. The detector also allows for energy calibration against the Michel edge or in $\pi^+\to e^+\nu_e$ decays without lowering the magnetic field. 


In conclusion, the HSRES technique significantly advances the experimental search for charged lepton flavor violation via the coherent muon-to-electron conversion channel. By circumventing the intrinsic limitations of conventional tracking approaches, namely material-induced electron scattering and energy losses, HSRES enables unprecedented improvements in electron energy resolution. This improved resolution directly enhances sensitivity to rare conversion signals, significantly extending the experimental reach beyond the Standard Model. Our results highlight the transformative potential of HSRES for precision exploration of fundamental symmetries.

\subsubsection{Acknowledgments}
\begin{acknowledgments}
The authors would like to thank the Project 8 collaboration for their inspiration, Elise Novitski for technical discussions regarding the CRES technique, and Marjorie Shapiro for inquiring whether the method could be applied to Mu2e and for stimulating this development. We thank Persephone Choi and Yuanyuan Ma for careful reading of the manuscript, and Jocelyn Luhr for the illustration of Fig.~\ref{fig:3d-detector}. We are indebted to the Mu2e collaboration for making the concept of late Vladimir Lobashev a reality and motivating us to pursue it further. This work was supported by the US Department of Energy (DOE) Office of High Energy Physics under Contract No. DE-SC0018988, and by the Physics Department at the University of California, Berkeley. This research used the resources of the National Energy Research Scientific Computing Center (NERSC), a
Department of Energy Office of Science User Facility, under NERSC award HEP-ERCAP0033800.
\end{acknowledgments}

Johan Vonk and Nicholas Cutsail contributed equally, with shared responsibility for simulation and data analysis.

\nocite{*}

\bibliographystyle{apsrev4-2}
\bibliography{google_scholar}

\end{document}